\RequirePackage{etoolbox}
\csdef{input@path}%
{%
 {sty/},% class file
 {img/},% graphics files
}
\documentclass[p04]{elsevierbook}
\usepackage{natbib}
\usepackage{lipsum}
\usepackage{booktabs}
\usepackage{url}
\usepackage{xcolor}
\usepackage{makecell}
\usepackage{multirow}
\usepackage{float}
\usepackage{subfig}

\definecolor{light-gray}{gray}{0.75}

\begin{document}

\Frontmatter

\Mainmatter
  \cleardoublepage

\begin{frontmatter}
\setcounter{chapter}{14}
\chapter{Volumetric video streaming}\label{chap2}
\subchapter{Current approaches and implementations}

\begin{aug}
\author%[addressrefs={ad1,ad2}]%
  {\fnm{Irene}   \snm{Viola}}%
  %{\fnm{Firstname}   \snm{Surname}\footnote{This is author footnote}}%
\author%[addressrefs={ad2}]%
 {\fnm{Pablo} \snm{Cesar}}%
\address%[id=ad1]%
  {Centrum Wiskunde en Informatica, Amsterdam, The Netherlands}%
% \address[id=ad2]%
%   {Long Address}%
\end{aug}

\begin{abstract}
    The rise of capturing systems for objects and scenes in 3D with increased fidelity and immersion has led to the popularity of volumetric video contents that can be seen from any position and angle in 6 degrees of freedom navigation. Such contents need large volumes of data to accurately represent the real world. Thus, novel optimization solutions and delivery systems are needed to enable volumetric video streaming over bandwidth-limited networks. In this chapter, we discuss theoretical approaches to volumetric video streaming optimization, through compression solutions, as well as network and user adaptation, for high-end and low-powered devices. Moreover, we present an overview of existing end-to-end systems, and we point to the future of volumetric video streaming.
\end{abstract}

% \minitoc

%
% \begin{chapterpoints}%[Chapter Points]
% \item The ends of words and sentences are marked by spaces. It doesn't
%   matter how many spaces you type; one is as good as 100.  The end of
%   a line counts as a space.

% \item The ends of words and sentences are marked by spaces. It doesn't
%   matter how many spaces you type; one is as good as 100.  The end of
%   a line counts as a space.
% \end{chapterpoints}

% \begin{dispquote}

%   The ends of words and sentences are marked by spaces. It doesn't
%   matter how many spaces you type; one is as good as 100.  The end of
%   a line counts as a space.

%   The ends of words and sentences are marked by spaces. It doesn't
%   matter how many spaces you type; one is as good as 100.  The end of
%   a line counts as a space.
  
%   \source{Name}

% \end{dispquote}

\end{frontmatter}

Since the first ACM Multimedia conference in 1993 \cite{multimedia1993}, video streaming over the Internet has been a major research topic for industry and academia. During the last 30 years the focus has shifted \cite{TOMM20years}: from the early video compression technologies from the late 1980s to media-focused streaming protocols in the 1990s, and from technologies for rate control and shaping based on Quality of Service (QoS) in the 2000s to dynamic Adaptive Streaming over HTTP and cloud rendering based on Quality of Experience (QoE) in the 2010s. The idea of Video-on-Demand, which was challenged after a number of unsuccessful trials by large media corporations \cite{Prashant2013}, have become an intrinsic part of our daily lives in the 2020s with even a Technology \& Engineering Emmy Award\footnote{https://theemmys.tv/tech-73rd-award-recipients/} in 2021 for the Standardization of HTTP Encapsulated Protocols. 

The focus of this chapter is on volumetric video streaming, which we anticipate will have a successful journey, even though bumpy and curved, ahead. In retrospective, we can see a number of similarities with video streaming from the 1990s, with promising and visionary services \cite{Yang2010, Kuster2012, Fuchs2014, orts-escolano_holoportation_2016}, some remarkable technological solutions \cite{alkhalili_survey_2020}, and upcoming standards \cite{schwarz2018emerging, graziosi2020}. Still, basic research is needed for ensuring the best possible, 6 Degrees of Freedom (6DoF), experience both for immersive consumption of media and for real-time communication. Some existing limitations include real-time compression and delivery techniques that are aware of the context and of the behaviour of the users, better modeling techniques of content that allow for dynamic media optimisation and tiling, and QoE-based systems that can accommodate to different environments and applications.

The chapter does not try to cover all aspects on immersive media technologies, since other chapters already provide an excellent overview on topics like compression and transmission of 360 videos and light fields. Other related technological areas like RGB-D \cite{dijkstra-soudarissanane_multi-sensor_2019, gunkel_vrcomm_2021} and free-viewpoint video \cite{Carballeira2021} systems, large-scale acquisition and storage systems \cite{Pauly2001, Golla2015, Tu2019}, and virtual worlds and environments \cite{Mondet2008, Lange2015} are outside the scope of this chapter.

The benefits of virtual reality, and volumetric video, are unquestionable, with the potential of radically transforming our lives \cite{Slater2016}. Already in the 2000s-2010s, significant effort went into 3D tele-immersion or virtual teleportation with initiatives like the Office of the Future\footnote{http://www.cs.unc.edu/Research/stc/index.html}, TEEVE \cite{Zhenyu2005}, and Viewport \cite{Zhang2013}. More recently, in 2014, the Moving Picture Experts Group (MPEG), started an ad-hoc group on point cloud compression, MPEG-PCC\footnote{https://mpeg-pcc.org}, where commercial solutions for this type of media mobilised research and industry towards a single direction. This chapter discusses the most recent different approaches and implementations on volumetric video streaming, in terms of media consumption and communication pipelines. Recent approaches include the reduction of the volume of data by removing redundancies and other non-noticeable aspects of media (e.g., occlusion based on field of view), the optimisation based on tiling and progressive streaming approaches combined with head-motion and movement prediction, and cloud and edge rendering of media for meeting the requirements of low-powered devices \cite{han_mobile_2019}.

\section{Theoretical approaches to volumetric video streaming}\label{sec2.1}
\begin{figure}
    \centering
    \includegraphics[width=\textwidth]{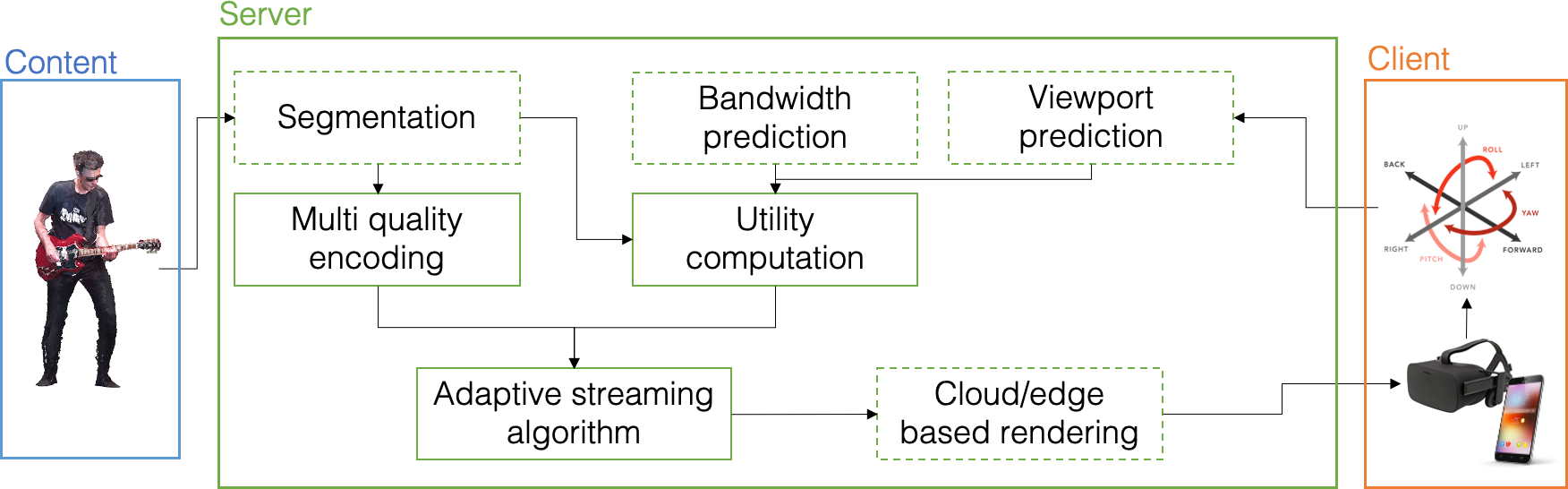}
    \caption{Diagram representing a generic server-client architecture for volumetric video streaming. The volumetric content is sent to the server, who is in charge of optionally segment it in partition, and then selecting the appropriate encoding for the client, given the network conditions and the user position and rotation. The content can then be streamed in volumetric form to the client, who will perform the rendering depending on the viewport; alternatively, the rendering can be offloaded to a cloud or edge server, and the viewport can be transmitted to the client as conventional 2D video. The model ``Electric guitar player'' is taken from~\cite{reimat2021cwipc}.}
    \label{fig:system}
\end{figure}

Volumetric media transmission involves large amounts of data in order to faithfully represent 3D objects and scenes, several orders of magnitude bigger than traditional images and videos (e.g., a point cloud video with around one million points requires 5Gbps). Thus, considerable effort has been spent in the literature to design, implement, and evaluate algorithmic solutions that would optimize transmission for the end user, limiting the bandwidth consumption without sacrificing the perceptual quality. In this regard, redundancies in the original data can be exploited to reduce the rate requirements. Moreover, several parts of the data might not be visible at any given time: for example, part of the object might be occluded (think of the back of a cube, which is not visible from the front), hidden by other 3D objects, or outside of the field of view. In this case, being able to predict, and exploit the position and field of view of the user that is visualizing the volumetric content can lead to sensible reductions in network expenditure, with little to no impact on the visual quality. It is no mystery, then, why user-adaptive strategies have become so popular for volumetric video streaming. An example of a generic volumetric video delivery system is depicted in Figure~\ref{fig:system}. The volumetric content is uploaded into a server, which is in charge of delivering it to the client at a given quality/bitrate level, depending on the network and device constraints. To aid in the delivery, several modules are available: the content might be segmented in order to exploit occlusions~\cite{petrangeli_dynamic_2019, park_volumetric_2018, park_rate-utility_2019, subramanyam_user_2020, van_der_hooft_towards_2019, liu_fuzzy_2020, liu_point_2021}; the viewport~\cite{han_mobile_2019, gul_kalman_2020} and the bandwidth~\cite{crowle_dynamic_2015, konstantoudakis_serverless_2021} might be predicted to facilitate delivery of upcoming packets; cloud- or edge-based rendering might be used to reduce strain on the client device~\cite{khan_can_2021, zhang_innovating_2021}.
An overview of challenges and opportunities for volumetric media streaming is presented by van der Hooft et al.~\cite{hooft_capturing_2020} and Liu et al.~\cite{liu_point_2021}.

Depending on the type of volumetric content representation that is adopted to transmit and render the data, different streaming strategies can be devised.  In the following, we detail streaming strategies for mesh and point cloud contents.%, as well as 2D video-based approaches. 

\subsection{Dynamic Mesh streaming strategies}

Early attempts at mesh streaming focused on using progressive encoding with several levels of detail to aid in transmission and reception in low-power devices. The seminal work of Hoppe et al.~\cite{hoppe_progressive_1996} introduces the concept of progressive meshes, which stores any arbitrary mesh $M$ as a coarse baseline $M_0$ along with $n$ detail layers that can be used to progressively increase the level of detail, up to the original mesh. The key concept at play in order to obtain the coarse baseline is the so-called \emph{edge collapse}: a transformation that unifies two adjacent vertexes into one. The inverse transformation, \emph{vertex split}, allows to reconstruct the original mesh from the coarser representation through each split record. The result is a lossless compression solution that achieves progressive decoding capabilities, at the expense of additional overhead. Further optimizations have been proposed since then, taking into account the progressive compression of attributes and the decoding time as additional factors. One of such optimizations is used as the basis for a web-based framework to stream meshes~\cite{lavoue_streaming_2013}, which allows for low-latency visualization with limited bandwidth. However, such scenario focuses on lossless delivery of static contents, mainly envisioned for scientific visualization. A strategy for optimizing the delivery of colored meshes in the event of packet loss is proposed by Cheng et al.~\cite{cheng2006packet}, in which the optimal subsampling factors for geometry and color are derived based on the network conditions; several overlapping subsampled versions of the mesh are created, put into packets, and sent to the receiving side in random order, in order to offer robustness against loss. 

In the case of dynamic sequences, it is common to differentiate between \emph{dynamic meshes} and \emph{Time Varying Meshes} (TMV). The first commonly refers to synthetic animated content, for which vertex and face count, as well as connectivity, remains constant; the varying element is the position of the vertexes. TVMs, inversely, 
% For dynamic sequences, the concept of Time Varying Meshes (TVM) was introduced to 
represent meshes with varying geometry, connectivity, vertex and face counts across frames, which commonly occurs when 3D acquisition of real-life sequences is performed. In this case, the lack of consistent correspondence between vertices across frames leads to complexity in handling the temporal redundancies. To date, several algorithms have been proposed to efficiently encode TVMs~\cite{gupta2002compression, gupta2003registration, han2008geometry, han2007time, doumanoglou2014toward}, the most popular in terms of ease of adoption being the open-source library Draco 3D Data Compression\footnote{https://google.github.io/draco/}, Corto\footnote{http://vcg.isti.cnr.it/corto/}, O3DGC\footnote{https://github.com/amd/rest3d/tree/master/server/o3dgc} and CTM\footnote{http://openctm.sourceforge.net/}. 
%benchmark
A benchmarking of open-source mesh codecs for interactive immersive media streaming is presented by Doumanoglou et al.~\cite{doumanoglou_benchmarking_2019}. The codecs are extensively compared based on bit-rate, distortion, and processing time, accounting for attribute and normals along with vertexes and connectivity. Results indicate that Draco and O3DGC are the best performing ones in terms of rate-distortion trade-off, whereas Corto offers the fastest decoding time. In addition, theoretical upper- and lower-bounds to the end-to-end latency are computed for each codec, along with an estimate of the achievable frame-rate, for several network conditions, exemplified by the Round Trip Time (RTT) values. Results indicate that, in case of small RTT, Corto is the best-performing one both in terms of latency and frame-rate. However, when larger RTT values are considered, Draco outperforms it in terms of latency.

Advanced compression algorithms significantly reduce the bandwidth expenditure needed to transmit meshes. However, to cope with the intense data requirements for low-latency streaming of mesh representations, adaptive solutions are often needed alongside efficient coding solutions. In the case of meshes, network monitoring and optimization has been successfully employed to reduce bandwidth requirements, for example, by designing a network adaptation service to monitor the network performance and consequently adapt the mesh compression parameters~\cite{crowle_dynamic_2015}. 
% Such an optimization is naturally a trade-off between rate and distortion, which is commonly measured using visual quality or Quality of Experience (QoE) metrics. 
%QoE
% 
%QoE optimization through deep learning
A network optimization system for a centralized immersive gaming setup that targets both the end user's QoE as well as the production costs for the transcoder is envisioned by Athanasoulis et al.~\cite{athanasoulis_optimizing_nodate}. The authors develop a cognitive network optimizer based on reinforcement learning, which monitors network metrics such as packet loss, bit- and frame-rate on the receiver side as well as in the transcoder, as well as the QoE of the transmitted meshes. The optimizer analyses the input and, if needed, instigates changes in the mesh compression level, both in the transcoder and receiver side, as well as redirecting the transcoder processing to either CPU or GPU. The optimizer is demonstrated through two profiles, based on whether it focuses on the QoE or it takes into account production costs, and compared with a baseline greedy approach. The results are evaluated based on the ratio between QoE and production costs, and show that considering both QoE and production cost leads to better performance with respect to the greedy approach in different bandwidth scenarios.
%serverless
Konstantoudakis et al~\cite{konstantoudakis_serverless_2021} propose a serverless framework for adaptive transcoding of meshes in a real-time immersive system. As part of the framework, a network optimization strategy to deal with the trade-off between QoE of each spectator and cost to the provider is devised. The model includes parameters such as the probability of a user to join or quit the transmission, the revenue for the provider for each user, the QoE for each user, and the costs of delivering and transcoding. Results demonstrate that the optimization can reduce the transcoding costs by $60\%$ and the delivery by $20\%$.

In the context of networking optimization, the choice of networking protocol is one key parameter for streaming of volumetric contents. Different protocols might offer more robustness at the expense of larger delays, or more agility when dealing with network changes. The impact of the quality of the mesh reconstruction, as well as the network delay, is studied in the context of a VR game experience with real-time acquisition and reconstruction~\cite{doumanoglou_quality_2018} with respect to the choice of network protocols. In particular, the authors 
% analyse how the resolution of geometry and texture affects the visual quality of the reconstructed players. Moreover, they 
investigate the impact of lag and frame drop on the final QoE, by selecting two different network protocols for the delivery, namely User Datagram Protocol (UDP) and Transmission Control Protocol (TCP). TCP relies on handshakes to ensure reliability and prevent packet losses. However, this comes at the expense of delays in the transmission chain. On the other hand, UDP is more agile, but does not have a recovery mechanism in place for lost packets; additionally, it does not have congestion control. Results of the experiment show that UDP was preferred to TPC, indicating that packet and frame loss is considered more acceptable with respect to large delays and latency.
% They also test two network protocols, namely UDP and TCP, to investigate the impact of lag and frame drop on the final QoE. Results indicate th
%DASH
The use of MPEG-DASH~\cite{sodagar2011mpeg} for adaptive streaming of 3D scenes is investigated by Zampoglou et al.~\cite{zampoglou_adaptive_2018}. In particular, they propose a framework to arrange X3D scenes in a Media Presentation Description (MPD), which will be used as the base element for the DASH architecture; similarly, they update the attributes in the Adaptation Set and Representation elements to be used for 3D objects while being compatible with the DASH architecture. The framework was evaluated in a prototype system against direct HTTP download, demonstrating how the DASH protocol was capable of delivering a first segment of the content, thus initiating the experience for the user, in a fraction of the time required to download the full content; moreover, the DASH delivery provided the full content in less time with respect to the HTTP counterpart.

%LOD for AR
Along with network adaptation, the user behaviour can be employed to optimize the delivery of meshes. A user adaptation strategy for multiple 3D objects in an AR scenario is proposed in~\cite{petrangeli_dynamic_2019}. The authors calculate the priority value of each object in the scene based on the user's field of view, using a method proposed by Chim et al.~\cite{chim2003cyberwalk}. Then, assuming that each object is available at several levels of detail, the contribution of each one to the final quality is computed, and a utility value is assigned to each object and level of detail, based on the ratio between quality and size. Finally, the adaptation strategy is formulated as a knapsack problem to maximize the number of objects (and relative levels of detail), based on their priority and utility.

% remote rendering
In order to reduce the amount of data to be sent to the client, cloud-based rendering has also been considered. In this case, the volumetric data is sent to an intermediate server, which renders and transmits the 2D view based on the user's head position and FOV. The approach has the advantage of avoiding the entire mesh to be transmitted to the receiver's side; however, this comes at the cost of increased latency, due to the necessity for the server to receive the user's position before rendering. To alleviate the problem, head motion prediction algorithms have been developed and tested for volumetric streaming. In~\cite{gul_kalman_2020}, a framework for head motion prediction based on Kalman filters is demonstrated. 
Khan et al.~\cite{khan_can_2021} investigate the use of several neural network architectures in order to achieve head movement prediction. 
Zhang et al.~\cite{zhang_innovating_2021} extend the problem to multi-user prediction, optimizing the transmission for both the QoE of the users and the network resource utilization. In particular, multicast is used to transmit frames for groups of users with similar viewports.%, in order to optimize network usage.

Immersive streaming applications bring a series of security concerns. In particular, when the 3D objects represent biometrics for identification such as human faces, there is need for secure solutions to maintain the privacy in a streaming scenario. In~\cite{tang_vvsec_2020}, a volumetric video attack is simulated, and a countermeasure based on adversarial perturbations is devised in order to dodge the attack without compromising the visual quality.

\subsection{Dynamic Point cloud streaming strategies}
Point cloud contents have recently seen a surge in popularity for volumetric video streaming scenarios involving natural acquisition. With respect to meshes, they have the advantage of being easier to process and manipulate, since each point can be treated independently as no connectivity information is required. However, in order to provide faithful representation of natural scenes, they require large collections of points to be delivered and rendered. Thus, first approaches in point cloud streaming focused on improving the compression efficiency for point cloud contents. 
%progressive
Similarly to what has been seen for mesh approaches, progressive point cloud encoders have been proposed to allow for refinement as the bandwidth increases, using octree structures that regularly partition the space\cite{peng2003octree, huang2006octree, schnabel2006octree}. Meng et al.~\cite{meng2003streaming} propose a progressive transmission method in which the level of details of the data are arranged hierarchically, so that the rendering can be performed faster in local areas defined by the users' field of view.
%I am not talking about the basic octree compression because I am assuming it is covered in another chapter; a reference can be added here.
%XOR streams
Kammerl et al.~\cite{kammerl_real-time_2012} present one of the first approaches for point cloud compression that is specifically tailored for online streaming. Specifically, they propose to leverage temporal redundancies by creating a double buffer octree structure to find correspondences between consecutive frames. The XOR operation is then applied to encode the differences between the frames. To be able to transmit leaf nodes with resolution greater than the octree resolution, they employ a point detail encoding module which transmits the difference between the leaf nodes and the voxel center. The same module is used to encode other attributes, such as texture and normals.
%MPEG Anchor
Mekuria et al.~\cite{mekuria_design_2017} present a compression solution for real-time encoding and decoding of colored point clouds. Their approach is based on an entropy-coded progressive octree structure with allows to select the appropriate level of detail. Rigid transform estimation is used to perform inter frame prediction, while the color attributes are encoded using JPEG. %They demonstrate their solution through a subjective experiment involving 20 test users.
Their solution was adopted as the reference encoder for the MPEG standardization efforts on point cloud compression~\cite{schwarz2018emerging}.

\begin{figure}
    \centering
    \includegraphics[width=\textwidth]{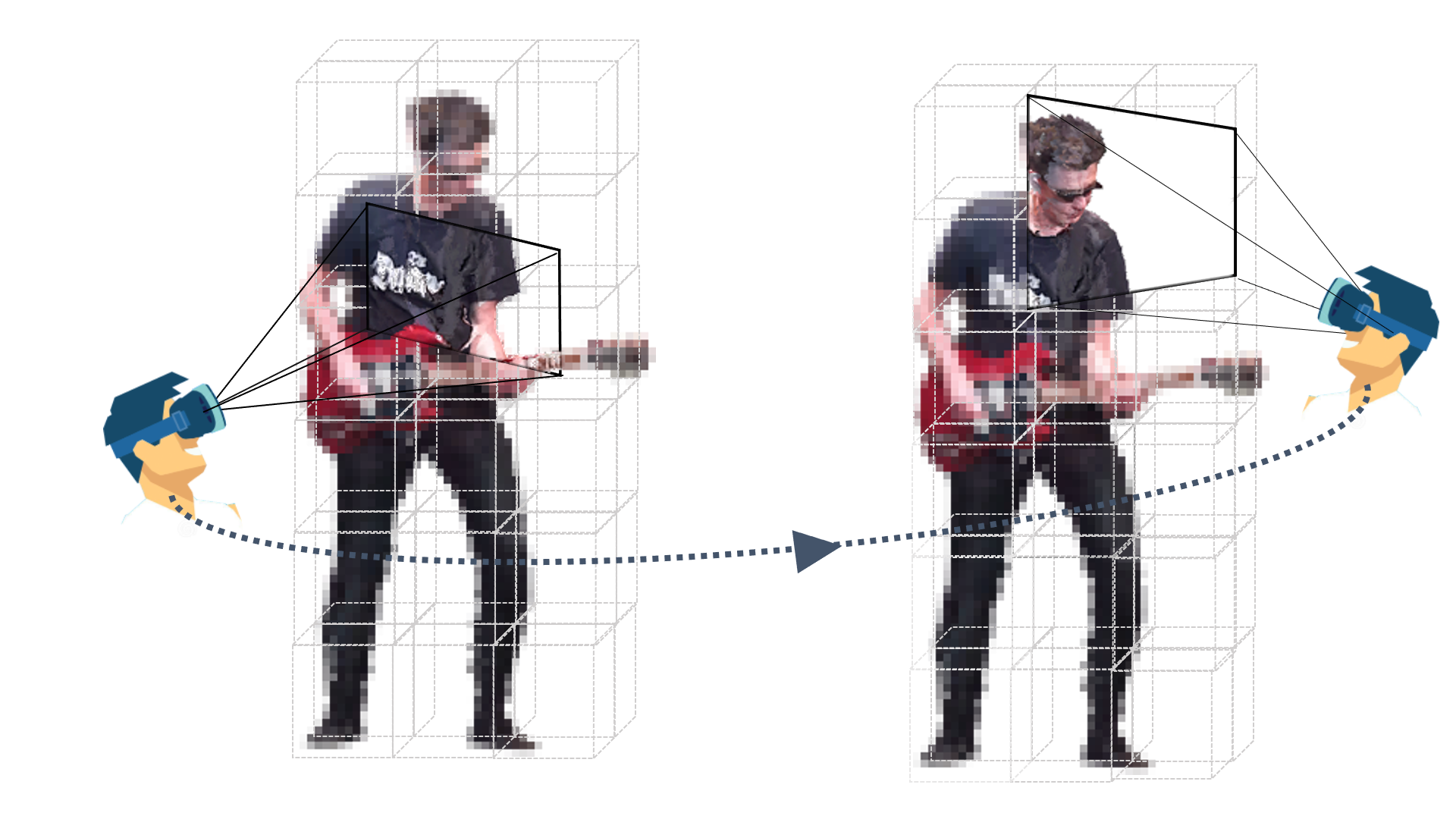}
    \caption{Example of user adaptive streaming. Depending on the user viewport, only the portion of the content that falls within the viewing frustrum is transmitted in high quality, whereas the remaining segments are transmitted in low quality. As the user changes position and orientation, different segments are selected. The model ``Electric guitar player'' is taken from~\cite{reimat2021cwipc}.}
    \label{fig:user_adaptation}
\end{figure}

%DASH-PC
Alongside efficient compression solution, adaptive streaming algorithms are needed to optimize delivery of point cloud contents over bandlimited networks. For point cloud contents, user adaptation represents one of the most adopted optimization solutions. As no connectivity information is needed to render them, segmenting the point cloud in non-overlapping regions to be encoded at different qualities is straightforward. An example of user adaptation is shown in Figure~\ref{fig:user_adaptation}. Only the regions that fall within the user viewport need to be transmitted and rendered in high quality, allowing for bandwidth savings while maintaining high visual quality. In order to allow for such optimization,
Hosseini et al.~\cite{hosseini_dynamic_2018} propose an adaptive streaming framework based on MPEG DASH, named DASH-PC. Multiple qualities of the same frame are created to form the \emph{adaptation set}, which is kept on an HTTP server. The client then requests the appropriate \emph{representation}, i.e., a frame at a certain quality level that complies with the bandwidth requirements. Each representation can be split into multiple \emph{segments}, which can help request only the parts of the content that are visible through the user's viewport. To create the representations, three subsampling algorithms are proposed; in addition, a human visual acuity model is employed to incorporate scaling into the adaptive algorithm, in order not to waste resources on details that would not be visible.
% 
%3D tiles
Park et al.~\cite{park_volumetric_2018, park_rate-utility_2019} propose a streaming setup for volumetric contents. They extend the concept of tiling, already used for adaptive streaming of omnidirectional contents, by introducing the concept of \emph{3D tiles}, and they define a utility function in order to estimate the importance of each tile for a given user. Their utility function takes into account the bandwidth cost of the selected representation, weighted by the number of distinguishable voxels in the tile, and the probability that such tile will be visible. They use a greedy maximization algorithm in order to select the tiles to be sent. Moreover, a window-based buffer is employed instead of a simple queue, in order to offer faster adaptation to user interaction.
%Shishir
A low-complexity tiling approach for real-time applications is proposed by Subramanyam et al.~\cite{subramanyam_user_2020}. Each point cloud is divided into non-overlapping tiles based on the visibility of each point by the camera that was used to acquire it, and the corresponding orientation of the camera is assigned to each tile. Then, each tile is compressed using a real-time encoder at different quality levels, forming the adaptation set. The utility of each tile is computed as a simple dot product between the orientation of the tile and the user's viewing angle. Three utility heuristics based on~\cite{van_der_hooft_towards_2019} are used to select the proper quality for each tile. The adaptive algorithm is tested under various fixed bitrate constraints, showing that adaptation leads to up to $60\%$ bitrate savings with respect to non-adaptive solutions. %QoE
Li et al.~\cite{li_qoe_2021} propose a QoE model to optimize volumetric video streaming. The model is based on the visual quality of the point cloud content, expressed through PSNR on both geometry and color; on the impairments deriving from stalling events, such as downloading and decoding time; and on the quality switch, which happens when a tile of different quality is requested.
%ViVo
A viewport prediction framework for mobile streaming of volumetric video is proposed by Han et al.\cite{han_vivo_2020}. The authors propose to segment the point cloud contents into cells, which are losslessly compressed using Draco. They employ a lightweight algorithm to perform viewport prediction, and they define three visibility-aware optimization models to select which cell to be transmitted to the user at each time segment: \emph{viewpoint visibility}, which considers an extended viewing frustum with varying level of detail; \emph{occlusion visibility}, which models whether the cell will be visible from the viewpoint; and \emph{distance visibility}, which uses objective quality measurements to understand which level of detail to assign to each cell. They demostrate their framework on 5G networks and on limited bandwidth scenarios, reporting significant gains on data usage and perceived quality with respect to the baseline.

%Multiple PCs
The previous algorithms tackled adaptation for single point cloud contents. Algorithms for multi point cloud rate adaptation are proposed by van der Hooft et al.~\cite{van_der_hooft_towards_2019}. In particular, the point clouds are ranked based on the distance with respect to the user, the visibility (and potential) of the point cloud, and the ratio between the visible area of the point cloud, and its bandwidth cost. Then, three utility maximization strategies are envisioned: \textit{greedy}, where the highest possible quality is given to the first ranked point cloud before moving down the rank; \textit{uniform}, where the bit budget is spent uniformly on all the point clouds, and the quality is increased one representation at the time; and \textit{hybrid}, where the uniform allocation is used for point clouds within the field of view, and any remaining budget is then used for point clouds outside of it. They test the impact of such heuristic considering different locations for the point clouds, as well as different camera paths to simulate user interactivity. Moreover, a QoE evaluation through subjective studies is performed in a subsequent work~\cite{van_der_hooft_objective_2020}, which demonstrates the significant impact of the bandwidth allocation strategy on the final perceived quality.

%Network optimization
The previous works have focused on providing user adaptation to cope with the bandwidth requirements of the systems. However, network optimization can play a large role in optimizing volumetric video streaming. Ramadan et al.~\cite{ramadan_case_2021} present an adaptive streaming mechanism specifically tailored for 5G networks. In particular, they propose adaptive content bursting in high bandwidth time windows, to ensure that the streaming can continue without stalling when low-bandwidth conditions occur. Moreover, they employ dynamic switching between 4G and 5G depending on the estimated channel conditions, to ensure a more stable streaming experience.
%fuzzy
Liu et al.~\cite{liu_fuzzy_2020, liu_point_2021} consider the impact of bandwidth changes, buffer status, and computational resources to design a fuzzy-based delivery system for tiled point clouds. The point cloud tiles are encoded at multiple quality level; however, one main difference with other approaches is that a coarse representation of the full point cloud is stored in the server, along with the \emph{decoded} version of the compressed tiles. The authors consider a quality maximization optimization problem in which both encoded and decoded versions of the tiles are available to be sent. The decision depends on the predicted bandwidth constraints, on the space availability on the buffer, and on the computational load, which are all modeled using fuzzy logic. They demonstrate gains with respect to the baseline using two point cloud models.

%ML models
More recently, some machine learning-based models have been employed to further enhance the capabilities of the volumetric video system, especially in adverse conditions, in order to deliver a better experience to the users. Zhang et al.~\cite{zhang_mobile_2020, zhang_efficient_2021} propose a super resolution-aided volumetric video streaming system. The proposed super resolution algorithm is optimized to reduce inference time, in order to cope with real-time streaming constraints, and is specifically designed to promote cross-frame consistency. The integration of the super resolution engine into the volumetric video streaming system is conducted through an adaptation model that takes into account the quality of the point cloud patches, the bandwidth consumption incurring from high resolution patches, and the computational resources needed to upsample the low resolution patches, along with the stalling that might derive from them.
Huang et al.~\cite{huang_aitransfer_2021} employ an end-to-end deep neural network which involves all the steps from acquisition to rendering and playback, thus avoiding traditional encoding, transmission, and decoding solutions. Key features are extracted from the input point cloud content and reconstructed at the receiving side using a lightweight neural network. An online adapter is added to switch between inference models depending on the bandwidth conditions. Their proposed system is validated through a real-time communication setup, in which contents are acquired by 3D sensors and transmitted through WiFi channels with varying bandwidth.

\section{Volumetric video streaming systems}\label{sec2.1}
In the previous section we have introduced some notable approaches to optimize transmission of volumetric contents for streaming purposes. Orchestrating such a system, however, is far from an easy feat. Technological limitations, network instability, system design constraints and device consumption costs are all aspects that need to be taken into consideration when constructing a feasible prototype for volumetric video streaming. Incorporating theoretical approaches may lead to the discovery of new vulnerabilities and hard constraints. In the following, we summarize some of the demonstration of volumetric video streaming systems, operating on meshes and point clouds.

\subsection{Mesh-based systems}

The majority of the systems proposed in the literature concerns 3D teleimmersion; as such, the focus in on creating a system that can acquire, process, transmit, and render volumetric objects in real-time. 
%describe teleimmersion?
A multi-camera system for 3D acquisition and transmission of real-world environments using meshes is described by Vasudevan et al.~\cite{vasudevan_high-quality_2011}. The objects are obtained from a cluster of calibrated cameras through disparity estimation; a coarse mesh model is derived through triangularization and then progressively refined through bisection. The bisection model is used for faster transmission; moreover, fast reconstruction is achieved through parallelization of data and rendering tasks. %to check
Mekuria et al.~\cite{mekuria_3d_2013} integrate efficient mesh compression and packet loss protection to their system. Their envisioned pipeline consists of a capturing module, an ad-hoc mesh compression module, a rateless coding module for packet loss protection, and a renderer. 
Beck et al.~\cite{beck_immersive_2013} present a group-to-group teleimmersive system. The participants are captured using multiple calibrated Kinect cameras; then, the data streams are processed using a parallelized and distributed processing pipeline. The rendering is achieved through projection-based multi-user 3D displays, which provides a perspective-corrected 3D scene visualization to each user. The system is evaluated in terms of usability in three scenarios: face-to-face meeting, side-by-side coupled navigation, and independent navigation.
Zioulis et al.~\cite{zioulis_3d_2016} also employ multiple Kinect cameras in order to obtain mesh models of their users. Their multi-camera setup captures the RGB-D frames asynchronously and sends them to a centralized server to convert to colored meshes; additionally, the user's motion is tracked through skeleton data. A server-based networking scheme is employed to transmit the 3D representations, which are compressed using static mesh encoders on a frame-by-frame basis. The system is demonstrated through a 3D gaming scenario.
Doumanoglou et al.~\cite{doumanoglou2018system} demonstrate a system architecture for an augmented virtuality scenario, in which user are 3D captured and can play a game in a teleimmersive system, using their body pose as a controller for the game. The system is comprised of a 3D capturing module with integrated pose recognition, player and spectator clients, and a 3D transcoder component to allow for adaptive streaming for both the players and the spectators. Furthermore, an adaptation for 5G networks is envisioned.

%AR
For a broadcasting scenario, a demonstration of a real-time volumetric streaming module for AR synchronized with broadcast video is given by Kawamura et al.~\cite{kawamura_real-time_2019}. They store the geometry in a binary file comprising of a list of vertexes and normals, to which the texture is attached after JPEG compression; bitrate savings are achieved by mesh simplification. The packets are sent over the internet using UDP, and synchronization between broadcast video and AR contents are achieved in their proposed receiver application through sync packets. %Their system transmits the uncompressed geometry as a binary list of verteces and normals, to with the texture is attached after being compressed with JPEG. 

% holoportation
Depending on the device that is used to experience the 3D contents, it might not be feasible to deliver the volumetric content directly to the end device. For example, unthetered devices might suffer from limited processing power, overheating issues, or battery limitations. Thus, it might be necessary to perform the rendering on an edge- or cloud-based server.
%holoportation
Orts-Escolano et al.~\cite{orts-escolano_holoportation_2016} design a volumetric telepresence system called ``Holoportation''. They use 8 near infrared cameras with active stereo depth estimation to capture the users; the conversion to meshes is done to ensure both spatial and temporal consistency, to improve the visual quality. Spatial audio is achieved by matching every user source to their relative 3D representation, along with spatialization. Lightweight compression is applied in order to comply with the real-time requirements. Finally, rendering is performed on the edge servers, employing head motion prediction to reduce latency, and the corresponding views are then transmitted to the rendering device. The system is extensively demonstrated in AR and VR scenarios.
%Cloud-based rendering
G\"{u}l et al.~\cite{gul_low-latency_2020, gul_low-latency_2020} propose a volumetric video streaming setup in which the meshes are sent to a cloud server, alongside the user position and rotation. The server performs 6DoF user movement prediction to forecast where the user will be and anticipate which rendered views to send to the user. The effect of the prediction module on the latency is evaluated with respect to the baseline, showing that the rendering error is reduced when employing the prediction mechanism.
An additional component that allows animation of volumetric data is designed for the system, which allows the volumetric video character to follow the user as it speaks or moves~\cite{gul_interactive_nodate, son_split_2020}.

\subsection{Point cloud-based systems}

\begin{figure}
     \centering 
     \subfloat[][Camera arrangement in the physical space (top view).]{\includegraphics[width=0.32\textwidth]{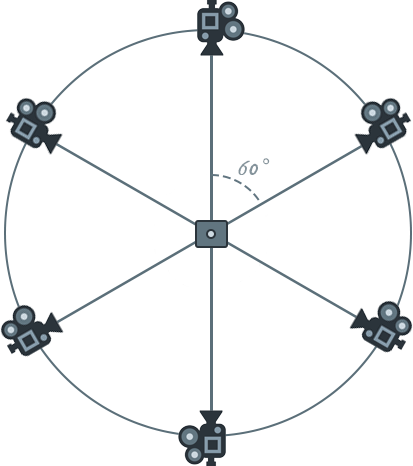}}\qquad \qquad
     \subfloat[][Example of user in the capturing setup.]{\includegraphics[width=0.47\textwidth]{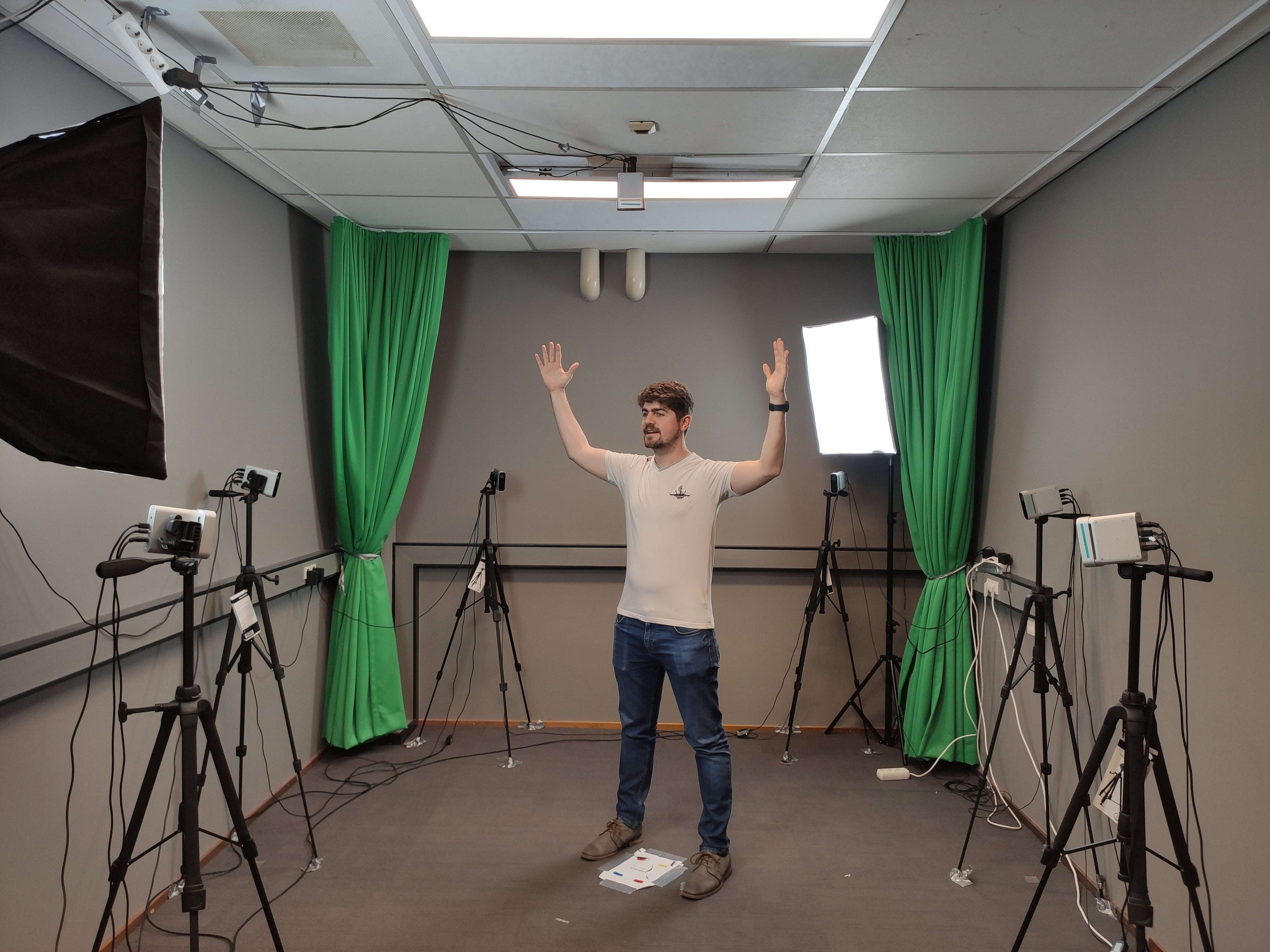}}
     \caption{Example of acquisition system for real-time transmission of volumetric media, using consumer-grade RGB-D sensors.}
     \label{fig:acquisition_system}
\end{figure}

A few systems have been proposed in the literature for point cloud video streaming; the main difference laying in whether they involve a real-time telepresence scenario~\cite{jansen_pipeline_2020, cernigliaro_pc-mcu_2020} or a broadcast volumetric video approach~\cite{qian_toward_2019, lee_groot_2020}.
%pipeline
Jansen et al.~\cite{jansen_pipeline_2020} adopt the DASH framework for real-time transmission of point cloud contents for 3D teleimmersion. Their proposed pipeline uses a multicamera Kinect setup (shown in Figure~\ref{fig:acquisition_system}) to acquire RGB-D data, which is converted to tiled point cloud representation at multiple quality levels. The data is sent to the DASH server, which is responsible for sending the selected quality to the receivers, which then decode and render the point cloud contents on the device.  
%pc-mcu
Cernigliaro et al.~\cite{cernigliaro_pc-mcu_2020} propose a Multipoint Control Unit (MCU) for real-time delivery optimization of multiple point cloud streams, called PC-MCU. The MCU server receives multiple point cloud representations, which are decoded and combined in the virtual scene. In order to optimize delivery for the receiving devices, the server considers the user's viewport and distance from the point cloud contents in order to lower the level of detail or remove contents outside of viewing window. The system is evaluated in terms of resource consumption, showing a reduction in both computational resources and necessary bandwidth, at the expense of added latency.

Special attention in the literature has been reserved for systems that are suitable for mobile phone consumption. With respect to tethered devices, there are some considerations that need to be kept in mind: the transmission should be optimized for wireless networks, to allow users to experience the content on the go and move around without constraints; the decoding time needs to be minimized not only for framerate requirements, but also for memory and power consumption, which are two critical features in mobile devices; the rendering needs to be optimized for fast user adaptation and low motion to photon latency, to maintain a pleasant user experience.
%nebula
Qian et al.~\cite{qian_toward_2019} present Nebula, a DASH-based system for volumetric streaming on mobile phones. In particular, they encode the point cloud content in layers, in order to achieve progressive streaming through DASH. The decoding and rendering is performed in an edge server to reduce the computational complexity for the mobile device. In order to reduce the latency, the edge server creates several rendered subframes, based on the current and predicted viewports, which are combined into a packet and sent to the mobile device. %The rate adaptation is performed based on a quality maximization principle.
%GROOT
Lee et al.~\cite{lee_groot_2020} present a system architecture for end-to-end streaming of volumetric video to mobile devices, named GROOT, powered by a faster, parallelized encoding and decoding scheme, along with viewport optimization such as frustum culling and depth-based sampling. Their encoding solution is based on octree geometry, but instead of encoding the occupancy of the entire tree, they split the octree structure at a predefined maximum breadth depth; then, the leaf nodes are individually encoded starting from their root at the maximum breadth depth to avoid dependencies with the branches. In practice, this allows to decode the desired leaf nodes without having to traverse the entire point cloud, facilitating faster rendering and removal of unnecessary points for visualization. The color compression efficiency is also improved by reordering the color information from the maximum breadth depth onward so to exploit spatial similarities. Frustum culling is applied by checking hierarchically whether the node boundaries fall within the frustum; for boundary cases, the children are checked until the entire boundary corners fall within. Finally, the sampling density is adjusted based on the perceived visual quality.

%
%AR remote rendering
% A remote rendering technique for real-time captured 3D meshes in the context of AR gaming is envisioned by Lee et al.~\cite{lee_implementation_2020}. The volumetric model is sent to the server, which performs a real-time parsing to 

% \begin{itemize}
    
    % \item Qian, F., Han, B., Pair, J., and Gopalakrishnan, V. (2019). Toward Practical Volumetric Video Streaming on Commodity Smartphones. Proceedings of the 20th International Workshop on Mobile Computing Systems and Applications, 135–140. https://doi.org/10.1145/3301293.3302358
    % \item Cernigliaro, G., Martos, M., Montagud, M., Ansari, A., and Fernandez, S. (2020). PC-MCU: Point cloud multipoint control unit for multi-user holoconferencing systems. Proceedings of the 30th ACM Workshop on Network and Operating Systems Support for Digital Audio and Video, 47–53. https://doi.org/10.1145/3386290.3396936
    % \item Jansen, J., Subramanyam, S., Bouqueau, R., Cernigliaro, G., Cabré, M. M., Pérez, F., and Cesar, P. (2020). A pipeline for multiparty volumetric video conferencing: Transmission of point clouds over low latency DASH. Proceedings of the 11th ACM Multimedia Systems Conference, 341–344. https://doi.org/10.1145/3339825.3393578
    % \item Lee, K., Yi, J., Lee, Y., Choi, S., and Kim, Y. M. (2020). GROOT: A real-time streaming system of high-fidelity volumetric videos. Proceedings of the 26th Annual International Conference on Mobile Computing and Networking, 1–14. https://doi.org/10.1145/3372224.3419214

% \end{itemize}

\section{Conclusion}
This chapter provides an overview of the current approaches of and implementations for volumetric video streaming, considering both media consumption and communication pipelines. Current solutions provide initial working systems that allow a first wave of novel applications from cultural heritage \cite{dwyer2021} to entertainment experiences \cite{Li2021}. Still, the possibilities are endless, from immersive performances \cite{Beacco2021} to future exhibitions and conference \cite{Ahn2021} to medical interventions \cite{riva2014}. Future work is needed for the development of such novel experiences, as well as the better understanding and modeling of the resulting content for optimisation purposes.

A major research area is the optimisation of the experience based on progressive streaming and tiling, currently inspired by previous work on 360 videos \cite{Zink2019, Fan2019}. Volumetric video, as a 6DoF experience, brings new challenges in terms of predicting and modeling the movement of the user and his/her relationship with the content. There are some initial investigations about \cite{Rossi2021}, but still more research needs to go in this direction. For example, there is a need for new datasets that focus on navigation patterns \cite{subramanyam_user_2020, zerman2021} for different contexts, which will help the development and validation of more advanced solutions.

Volumetric video will allow for interactive and immersive experiences, palliating existing problems like the Zoom fatigue \cite{Bailenson2021}. Future research is needed for the provision of adequate interaction mechanisms and the seamless inclusion of interactive content in the experiences. While recently works are focusing on trying to better understand the basic constructs on social VR \cite{Yassien2020, Williamson2021}, further research is needed for considering the volumetric video case. Finally, more datasets centered in interactive activities are missing \cite{reimat2021cwipc}.

Networks, in particular mobile, are evolving, bringing closer to reality futuristic scenarios. Edge rendering, for example, is helping to make volumetric video available in mobile devices. New protocols and infrastructure are coming, increasing the opportunities for everywhere anytime volumetric video consumption. For example, the European Commission has recently launched an initiative on Smart Networks and Services (SNS)\footnote{https://digital-strategy.ec.europa.eu/en/policies/smart-networks-and-services-joint-undertaking} towards 6G, where eXtended Reality will become a core use case. 

\bibliographystyle{elsarticle-num}
\bibliography{bibl}

% \lipsum[1-8]

%   \include{part2_omnidirectional}
%   \include{chapter1}
%   \include{chapter2}
  
%   \include{part3_lightFields}
%   \include{chapter1}
%   \include{chapter2}
  
%   \include{part4_volumetric}
%   \include{chapter1}
%   \include{chapter2}
  
%   \include{part5_applications}
%   \include{chapter1}
%   \include{chapter2}

%   \appendix
%   \include{appendix01}
%   \include{appendix02}

\Backmatter

\end{document}